\magnification=\magstep1
\tolerance=500
\vskip 2true cm
\rightline{TAUP 2701-2002}
\rightline{24 June, 2003}
\bigskip
\centerline{\bf Eikonal Approximation to 5D Wave Equations}
\centerline{\bf as}
\centerline{\bf  Geodesic Motion in a Curved 4D Spacetime}
\bigskip
\centerline{O. Oron and L.P. Horwitz}
\smallskip
\centerline{School of Physics and Astronomy}
\centerline{Raymond and Beverly Sackler Faculty of Exact Sciences}
\centerline{Tel Aviv University, Ramat Aviv 69978, Israel
\footnote{*}{Also  at Department of
Physics, Bar Ilan University,  Ramat Gan 529000, Israel}}

\bigskip
\noindent {\it Abstract:\/}We first derive the relation between the
 eikonal approximation to the Maxwell wave equations in an inhomogeneous
anisotropic medium and geodesic motion in a three dimensional
Riemannian
manifold using a method which identifies the symplectic structure of
the corresponding mechanics.  We then apply an analogous method to the
five dimensional generalization of Maxwell theory required by the
gauge invariance of Stueckelberg's covariant classical and quantum
dynamics to demonstrate, in the eikonal approximation, the existence
of geodesic motion for the flow
of mass in a four dimensional pseudo-Riemannian 
manifold. No motion of the medium is required. These results provide a
 foundation for the 
geometrical optics of the five dimensional radiation theory and
establish a model in which there is mass flow along
geodesics. Finally, we  discuss the interesting case of relativistic
quantum theory in an anisotropic medium as well.  In this case the
eikonal approximation to the relativistic quantum mechanical current
 coincides with  the 
geodesic flow governed by the pseudo-Riemannian metric obtained from the
 eikonal approximation to solutions of the Stueckelberg-Schr\"odinger
equation.
 This construction provides a model for an
underlying quantum mechanical structure for classical dynamical
motion along geodesics on a pseudo-Riemannian manifold. The
locally symplectic structure which emerges is that of Stueckelberg's
covariant mechanics on this manifold.
\vfill
\break
\bigskip
\noindent
{\bf 1.  Introduction}
\smallskip 
\par It has been known for many years that the Hamilton-Jacobi
equation of classical mechanics
defines a function which appears to be the eikonal of a wave equation,
and therefore that classical mechanics appears to be a ray
approximation to some wave theory$^1$. The propagation of rays of
waves in inhomogeneous media appears, from this point of view (as a
result of the application of Fermat's principle), to correspond to
geodesic motion in a metric derived from the properties of the
medium$^2$.   This geometrical
interpretation has been exploited recently by several authors to
construct models which exhibit three dimensional analogs of general
relativity by studying the wave equations of light in an inhomogeneous
medium$^3$, and, to achieve four dimensional analogs,
 sound waves and electromagnetic propagation in
 inhomogeneously moving materials$^4$.  Visser {\it et al\/}$^{5,6,7}$
have pointed out that condensed matter systems such as acoustics in flowing
fluids, light in moving dielectrics, and quasiparticles in a moving
superfluid can be used to mimic kinematical aspects of general
relativity. Leonhardt and Piwnicki$^8$ and Lorenci and Klippert$^9$,
for example, have discussed the case of electromagnetic propagation in
moving media.   In order to  achieve four dimensional geodesic flows, it has
been necessary to introduce a motion of the medium. 
\footnote{*}{We remark that Obukhov and Hehl$^{10}$ have shown that a conformal class
of metrics for
spacetime can be derived by imposing constrained linear constitutive
relations between the electromagnetic fields $(E,B)$ and the
excitations $(D,H)$, using Urbantke's formulas$^{11}$, developed to
define locally integrable parallel transport orbits in Yang-Mills
theories (on tangent 2-plane elements on which the Yang-Mills
curvature vanishes).}
  There is considerable interest in extending these
analog models for the kinematical aspects of gravity to include
dynamical aspects, {\it i.e.}, considering gravity as an emergent 
phenomenon$^{5,6}$.
\par The manifestly covariant classical and quantum mechanics
introduced by \hfil\break
Stueckelberg$^{12}$ in 1941 has the structure of
Hamiltonian  dynamics with the Euclidean three dimensional space replaced
by four-dimensional Minkowski space (since all four of the components
of energy-momentum are kinematically independent, the theory is
intrinsically ``off-shell'').  The dynamical evolution of the system
is governed by a ``world time'' $\tau$. This theory leads to
 five dimensional
wave equations for the associated gauge fields. The
self-interaction problem of the relativistic charged particle has
been recently studied$^{13}$, where it was shown that the radiation
field is associated with excursions from the mass shell; highly
non-linear terms appear in the resulting generalized Lorentz force.
In the limit that the motion maintains a state very close to the mass
shell limit, the equation reduces to that of Dirac$^{14}$.
\par The known bound state spectra and corresponding wave functions
for the (spinless)
two-body problem in potential theories formulated in a manifestly
covariant way have been worked out$^{15}$. The classical relativistic
Kepler orbits have been studied in detail$^{12,16}$. 
 \par In order for the Stueckelberg-Schr\"odinger
equation of the quantum form of this theory to be gauge invariant, it
is necessary to introduce a fifth gauge field, compensating for the
derivative with respect to the invariant evolution parameter $\tau$ $^{17,18}$.
Generalized gauge invariant field strengths, $f_{q,p}$, with $q,p
= 0,1,2,3,5$ occurring in the Lagrangian to second order generate
second order field equations analogous to the usual Maxwell equations,
with source given by the four-vector matter field current and an
additional Lorentz scalar density.  The canonical second quantization
of this theory was studied in ref. $19$.
 Taking the Fourier transform of
these equations over the invariant parameter, as we demonstrate below,
one seees that the zero
frequency component (zero mode) of the equations coincides with the
standard Maxwell theory (the fifth field decouples).
The Maxwell theory is therefore properly contained in the
five-dimensional generalization as the zero frequency component. 
 The general form of the theory has variations as a
function of world time, about the (world time independent) Maxwell theory.
 
\par In the quantum
case the four-currents are given by bilinears in the wave
function containing first derivatives, and the fifth source is the
(Lorentz) scalar probability density. The symmetry of the
homogeneous equations, which can be $O(3,2)$ or $O(4,1)$, depending on
the sign chosen for raising and lowering the fifth index; it is not
realized in the inhomogeneous equations, since the spacetime current
components and the fifth component (the quantum mechanical probability
density $|\psi|^2$) cannot be transformed into each other by a linear
transformation.  Hence, without augmenting
the symmetry of the matter fields beyond $O(3,1)$, the
fifth field, whose source in the Maxwell-like equations is the
probability density (or, classically,the matter density) in
spacetime, can play a special role. There appears to be no kinematic
basis for choosing one or the other of these signatures; atomic
radiative decay, for example, contains points in phase space (for
radiation of off-shell photons) for either type.  We note, however,
that the homogeneous equations corresponding to the $O(4,1)$ signature
appear, under Fouries transform of the $\tau$ variable, as 
Klein-Gordon type wave equations with positive mass-squared (physical
particles) , but for the $O(3,2)$ choice of signature, these equations
have the ``wrong'' sign (tachyonic) for interpreting the additive term
as a mass-squared. As a
physical example of how both metrics may play a role, let us suppose that
  the off-shell radiation impinging on an atomic nucleus, for
example, has the $O(3,2)$ signature.  During the interaction,
as the state evolves as a function of $\tau$,
the metric may make a transition to the $O(4,1)$ form. This transition could
correspond to the observed phenomenon of photoproduction (the vector
dominance model). We
therefore leave open the question of a definitive choice of the
signature for the five dimensional radiation field at this stage.
\par In either case,  the 
four dimensional spacetime submanifold coincides with that of 
Minkowski; on this manifold, the fields are defined as a family of 
functions paramaterized by  $\tau$.  This 
parameter controls the evolution generated by the Hamiltonian of the 
system of particles and radiation. That this parameter occurs in the wave 
equations with timelike or spacelike metric does not change its physical 
interpretation.  At each value of $\tau$, the four dimensional 
configuration corresponds to a Maxwell wave; these waves, however,
evolve non-trivially in $\tau$, reflecting the dynamics of the
system. For example, in classical charged particle scattering,
 the conserved
currents are constructed (as we remark below) in the four-dimensional
Maxwell theory, by integration of a local current ${p^\mu \over
M}\delta^4(x-x(s))$ over all proper time $s$ (which we may identify
here with the world time). The trajectory $x^\mu(s)$ is not, however,
known until the problem is solved; the system is not {\it a priori}
integrable. In the higher dimensional theory we discuss here,
equations of motion determine the trajectory $x(\tau)$, and in the
neighborhood of a given $\tau$, the generalized (sometimes called
pre-Maxwell) equations determine the fields as functions of
$\tau$. This system is explicitly, in principle, integrable. The
integral of the results over all $\tau$ (on both fields and currents) 
then provides exact solutions of
the corresponding Maxwell problem, with currents computed from the
resulting trajectories in the standard way.  Since, however, the
Lorentz force
is non-linear, it is clear that the dynamics forming the trajectories
may be very different from that of the usual Lorentz force$^{13}$. 
 \par The structure of the gauge theory obtained from the
non-relativistic Schr\"odinger equation is precisely analogous.  The
four dimensional gauge invariant field strengths obey second order
equations, for which the sources are the vector currents and the scalar
probability density (or, classically, the matter density in three
dimensional space). No linear transformation can connect the
Schr\"odinger probability density with the vector currents, so that
the scalar density and the fourth gauge field can play a special role.
\par  Since the eikonal approximation naturally lowers the dimension
of the differential equations describing the fields by one, the
eikonal approximation to the five-dimensional field equations results
in four dimensional differential equations.  In the presence of a
non-trivial dielectric structure of the medium, the four dimensional
field equations resulting from  the eikonal approximation can describe
 geodesic motion in four dimensional
spacetime without the necessity of adding motion to the medium. We
emphasize that the underlying manifold, on which the fields are
defined, is a flat Cartesian space, but that the
dynamically induced trajectories are curved, and can be described by
the geodesics of a pseudo-Riemannian manifold. This
result forms our basic motivation for studying the generalized
dynamics of Stueckelberg$^{12}$ in this context. 
\par We start with a review of the simpler case of Maxwell wave 
propagation
in a three dimensional inhomogeneous, anisotropic medium in which we
introduce a method which permits us to identify clearly the structure
of the effective mechanics which emerges in the eikonal approximation.
 In Section 3, we study  
the eikonal structure of waves in a 5D inhomogeneous medium, in which
 Minkowski spacetime is embedded, and in Section 4,  a wave equation
 of Schr\"odinger type, and show that the resulting rays 
have a precise analog with the results of Kline and Kay$^2$ (who have
studied the three dimensional ray limit of the usual Maxwell
equations), but in a
 4D spacetime manifold with a pseudo-Riemannian structure. Kline and
Kay$^2$ show that the rays are geodesic in the metric associated with
the anisotropic inhomogeneous medium.  As for the case of Kline and
Kay$^2$, it follows from the existence of a Hamiltonian that the
corresponding Lagrangian obeys an extremum condition which describes
the rays as geodesics.  We show
that there is mass flow along these rays, and that
the flow is controlled by generating functions of Hamiltonian type,
establishing a relation between geodesic flow and a relativistic
particle mechanics of symplectic form.
\bigskip
\noindent
{\bf 2. Eikonals of the Maxwell Wave Equation}
\smallskip
\par We first consider the case of Maxwell wave propagation
in a three dimensional inhomogeneous, anisotropic dielectric media,
as in ref. 2. We
find the Fresnel surfaces in terms of quadratic forms which are the
solutions of an eigenvalue
equation, where each polarization is associated with a Riemannian
metric$^{2,20}$ (of Finsler type) giving rise to geodesic flow.  The
 study of the eigenvalue equations, rather than the determinant
 condition used in ref. 2, permits us to identify more clearly the
 structure of the effective mechanics which emerges.
\par We start by writing Maxwell's equations in
an inhomogeneous, anisotropic medium.   These are of the usual form
$$ \nabla\cdot{\bf D}=\rho \eqno(1a)$$
$$\nabla \times {\bf H}-{\partial {\bf D}\over
 \partial t}={\bf J} \eqno(1b)$$
$$ \nabla\cdot{ \bf B} = 0 \eqno(1c)$$
$$ \nabla \times {\bf E}+{\partial {\bf B}\over \partial t}=0,
\eqno(1d)$$
where $\rho$ is the charge density, ${\bf J}$ the current density of
the sources, and, with $i,j=1,2,3$ (we assume here that the
coordinates labelled by these indices are Cartesian, and we write
indices up or down for convenience of expression), where, however, the relations
$$D_i=\epsilon_{ij}({\bf x}) E_j \,\,\, , \,\,\, B_i=\mu_{ij}({\bf x})H_j \eqno(2)$$
reflect the properties of the medium (we assume all medium properties
to be independent of time and that, for our present purposes, there is no
mixing between electric and magnetic fields through the constitutive
tensor relations). We then  multiply Eq.$(1d)$ with 
the matrix
$\mu^{-1}$, act with the $curl$ operator and substitute Eq.(1b). We
obtain, in the absence of sources,
$$ { \nabla}\times (\mu^{-1} ({ \nabla} \times {\bf E}))+{\partial^2
{\bf D}
 \over \partial t^2}=0 \eqno(3)$$
 Substituting for $\bf D$ using Eq.$(2)$, one obtains
$$\epsilon^{-1}({ \nabla}\times (\mu^{-1} ({ \nabla} \times
 {\bf E})))+{\partial^2 {\bf E} \over \partial t^2}=0 \eqno(4)$$
We rewrite Eq. (4) in  index notation (we use Latin indices for space
components $1,2,3$, and $t$ for the time index):
$$(\epsilon^{-1})^{rl}\varepsilon^{lmn}
{\partial}^m(\mu^{-1})^{ni}(\varepsilon^{ijk}
{\partial}^jE^k)+{\partial}^2_t E^r=0 \eqno (5)$$
Assuming a solution of the form ${\bf E} = {\bf A} \exp{i\omega
(t-\varphi({\bf x)})} $ and using the eikonal
 approximation (for large $\omega$), one obtains
   $$ -(\epsilon^{-1})^{rl}\varepsilon^{lmn} (\mu^{-1})^{ni}
 \varepsilon^{ijk}{\partial}^m \varphi{\partial}^j \varphi A^k= A^r. 
\eqno (6)$$
\par For a given wave front at a given time $\varphi ({\bf x})
{\vert}_{t=t_0}=c_0$ one can think of curves, originating from each
point on the wave front, and everywhere in the direction of the gradient of
$\varphi(\bf x)$. These lines can be considered to be trajectories on
which the different elements of the wave-front propagate. The
trajectories of the wave-front normals, in an anisotropic medium,
however, are
in general not in the same direction at a given point as the direction
of propagation of the energy of radiation, i.e., of the Poynting
vector. We shall
refer to the trajectories which follow the {\it Poynting vectors} as 
{\it rays}. We now show that
the equation for the rays is a geodesic equation. The metric
determining the geodesics is fixed by the electric and
magnetic properties of the medium and, in general, depends on the
polarization of the field.  
\par We denote the wave front gradient (which can be interpreted, as
will be seen later, as the momentum flowing along the ray)
$$p_m={\partial}_m \varphi.$$ 
Equation (6) then takes the following form:
$$ - (\epsilon^{-1})^{rl}\varepsilon^{lmn} (\mu^{-1})^{ni}
 \varepsilon^{ijk}{p}^m{p}^j A^k= A^r, \eqno (7)$$
Multiplying Eq. $(7)$ by ${p}^s \epsilon^{sr}$, the left hand
side vanishes, and we obtain
$$ {\bf p}\cdot (\epsilon {\bf A})=0, \eqno (8)$$
the eikonal form of Eq.$(1a)$.  The eikonal solution is therefore consistent 
with all of Maxwell equations in the non-homogeneous medium.
We now define (obviously positive definite for $\epsilon$ and $\mu$
scalar; the symmetric in $mj$ part is symmetric in $rk$ if $\epsilon$
and $\mu$ are symmetric)
$$ {M}_{mj}^{rk} = -(\epsilon^{-1})^{rl}\varepsilon^{lmn}
(\mu^{-1})^{ni} \varepsilon^{ijk} \eqno(9) $$
\par   Due to
the relation $(8)$, the $3 \times 3$  matrix $M_{ij}^{rk}p_i p_j $ acts
 in a two dimensional subspace; one may replace ${\bf A}$ by
 $ v=\epsilon{\bf A}$ if ${M}_{mj}^{rk} $ is replaced by $\epsilon_{ml}
 {M}_{ij}^{lm} \epsilon^{-1}_{mk}$,
 which takes the two dimensional subspace orthogonal to ${\bf p}$ into itself.
Eq.$(7)$ may then be written as
$$ {\hat M}_{ij}^{rk}p^i p^j v^k = v^r ,\eqno(10)$$
where ${\hat M} = \epsilon M \epsilon^{-1}$. Eq.$(10)$ clearly imposes
 a condition on the magnitude of the vector ${\bf p}$. As we shall see,
it is restricted to two discrete values.
\par If we express the tensor $ {\hat M}_{mj}^{rk} $ in coordinates for
which the 3 direction is parallel to the momentum ${\bf p}$, we obtain
the eigenvalue condition
$$ {\hat M}'^{pq}_{33} p_3'^2 v'^q = v'^p, \eqno (11)$$
where the primed quantities are in the momentum oriented coordinate
system; since $v'$ provides support only in the two dimensional
subspace orthogonal to ${\bf p}$, the matrix $ {\hat M}'^{pq}_{33}$
is nonzero only in that subspace.  It is not, in general, degenerate.
If the eigenvalues are $\lambda_{(\alpha)},\, \alpha =1,2$, it follows
from $(11)$ that $p_3'^2 = \vert{\bf p}\vert^2 $ must have
the values $\vert{\bf p^{(\alpha)}}\vert^2=
\lambda_{(\alpha)}^{-1}$.  Transforming back to the
original frame, we see that $(10)$ can be satisfied only for two
polarization modes $v^k_{(\alpha)}$, and corresponding momenta of the
same direction with
magnitudes $\vert {\bf p}^{(\alpha)}\vert^2 =
\lambda_{(\alpha)}^{-1}$.
 Multiplying, for each $\alpha$,  both sides by (normalized)
 $ v_{(\alpha)}^r$ and summing on $r$, we obtain the form
$$ H^{(\alpha)}= p_i^{(\alpha)} p_j^{(\alpha)} g^{(\alpha)}_{ij}- 1 =0,
 \eqno(12)$$
 where
$$g^{(\alpha)}_{ij}= v_{(\alpha)}^r{\hat M}_{ij}^{rk}
v_{(\alpha)}^k. \eqno(13)$$
  We shall identify ${\bf p}^{(\alpha)}$ below as a canonical
momentum, orthogonal to
the surface defined by $\varphi$;  the (${\bf p}^{(\alpha)}$
 direction dependent) matrices
$ g^{(\alpha)}_{ij}$ therefore act as (Finsler type) metrics for each
of the polarizations$^{20}$.
\par Since $g^{(\alpha)}$ depends only on ${\bf x}$ and the direction of 
${\bf p^{(\alpha)}}$
 at any given point in space, the condition $(12)$ determines
the magnitude  of ${\bf p}^{(\alpha)}$, and therefore describes a surface. These
surfaces, described for $\alpha =1,2$, coincide with the Fresnel
surfaces defined by Kline and Kay$^2$ by means of the determinant of
coefficients of the eigenvalue equation. We have examined here directly the
 eigenvalue equations since we shall follow this method in the 5D
 case.  
It is shown for the three dimensional case in ref. 2 (the proof is
given below for the similar four dimensional problem) that
${\partial}_{p^{(\alpha)}_i} H^{(\alpha)}$ is
parallel to the
Poynting vector (clearly the same direction for each $\alpha$).
If we parametrize the flow along a given ray at ${\bf x}$ with some
parameter $s$, this statement can be written as
$${\dot x}^i={ dx^i \over ds}=\lambda \partial_{p^{(\alpha}_i}
 H^{(\alpha)}, \eqno(14)$$
where  $\lambda$ is a scale on $s$. 
The total derivative  of $H^{(\alpha)}$ with respect to $x_i$ is given by 
$${dH^{(\alpha)} \over dx^i}={\partial H^{(\alpha)} \over \partial x^i}+
 {\partial H^{(\alpha)}
\over \partial p^{(\alpha)}_k} {\partial p^{(\alpha)}_k \over  
\partial x^i}. \eqno(15)$$
This quantity must vanish, since the derivative relates neighboring
Fresnel surfaces, on which (in this mode) $H^{(\alpha)}$ is zero.
Substituting $(14)$ in $(15)$ and using
$${ \partial p^{(\alpha)}_k \over \partial x^i}={\partial^2 \varphi
\over
 \partial x^i \partial x^k }={\partial p^{(\alpha)}_i \over  
\partial x^k}, \eqno(16)$$
we obtain
$$\lambda {\partial H^{(\alpha)} \over \partial x^i}+ {\dot x}^k
 {\partial p^{(\alpha)}_i \over  \partial x^k} =0 \eqno(17)$$
which gives 
$${\dot p}^{(\alpha)}_i=-\lambda \partial_{x_i} H^{(\alpha)} . \eqno(18)$$
Eqs. $(14)$ and $(18)$ correspond to the locally symplectic structure of a
Hamiltonian flow generated by the function $(12)$ in each mode.
Moreover, one sees that the geodesic equation associated with the
metric $g^{(\alpha)}$ is equivalent to this Hamiltonian flow. This result
agrees with application of the Fermat principle. 
\bigskip
\noindent
{\bf 3. Eikonals of the 5D Wave Equation}
\smallskip
\par In this section we apply a technique similar to that used above
to study
 the structure of wave equations in five dimensions which
follow as a consequence of the requirement of gauge invariance of the 
covariant classical
and quantum mechanics of Stueckelberg$^{12, 17-19,21}$.  We show that
there is a Hamiltonian form for the generation of the rays in this
case as well, and that these rays form spacetime geodesics in a metric
 space determined by the ``dielectric'' constitutive tensor relating
 the tensor fields (analogous to $(E, B)$) and their corresponding 
excitation fields (analogous to $(D, H)$).
\par  We shall use
Greek letters for  space time indices
 ($\mu=0,1,2,3$) and Latin letters to include
a fifth index representing the Poincar\'e invariant $\tau$ parameter in
addition to the usual 4 spacetime coordinates (e.g., $q=0,1,2,3,5$). 
The analysis proceeds by a generalization of the method discussed in
 Section 2 above.
The generalized electromagnetic field tensor is  written
$$f_{q_1\, q_2} \equiv {\partial}_{q_1}a_{q_2}-{\partial}_{q_2}a_{q_1},  $$
where $a_q$ are the so-called pre-Maxwell electromagnetic
potentials ( the fifth gauge potential $a_5$ is
required for gauge
compensation of $i\partial_5$, generating the  evolution of the 
Stueckelberg wave function$^{12,17}$).
\par We introduce the dual (third rank) tensor
$$ k^{l_1 \, l_2 \, l_3}=\varepsilon^{\, l_1 \, l_2 \, l_3 \, q_1 \,
q_2}f_{\, q_1 \, q_2},$$
where $\varepsilon^{\, l_1 \, l_2 \, l_3 \, q_1 \,
q_2}$ is the antisymmetric fifth rank Levi-Civita tensor density.
The homogeneous pre-Maxwell equations are then given by
$${\partial}_{l_3}k^{\, l_1 \, l_2 \, l_3}=0, \eqno(19)$$
or, more explicitly (the 5 index is raised with signature $\pm$,
according to, as discussed above, $O(4,1)$
or $O(3,2)$ symmetry of the homogeneous field equations),
$${\partial}_5 \varepsilon^{l_1 \, l_2 \, 5 \, q_1 \, q_2}f_{q_1 \, q_2}+{\partial}_{\sigma} \varepsilon^{l_1 \, l_2 \,\sigma \, q_1 \, q_2 }f_{q_1 \, q_2}=0 . \eqno(20)$$ 
We now divide Eq.$(20)$ into two cases. In the first, we take the 
indices $l_1, l_2$ to  correspond only to space-time indices: 
$$ {\partial}_{5} \varepsilon^{\mu \nu \, 5 \lambda \sigma
}f_{\lambda \sigma }+2{\partial}_{\sigma} \varepsilon^{\mu \nu \sigma
\lambda 5}f_{\lambda 5}=0,$$
or
$${\partial}_{5}
\varepsilon^{\mu \nu \,  \lambda \sigma }f_{\lambda \sigma
}+2{\partial}_{\sigma} \varepsilon^{\mu \nu \sigma \lambda }f_{\lambda
5}=0, \eqno(21) $$
where $\varepsilon^{\mu\nu\sigma\lambda}$ is the four dimensional
Levi-Civita tensor density.  
This equation, on the 0-mode ($\tau$ independent Fourier components)
does not involve any of the usual Maxwell fields but only the fifth
(Lorentz scalar) electromagnetic field. The second set from Eq.$(20)$
corresponds to $l_1 \, {\rm or}\,  l_2=5 $. It is clear then that all the
 other 4-remaining indices must be space-time indices and we obtain
$$ \varepsilon^{5 \mu \sigma \delta \nu}{\partial}_{\sigma} f_{\delta
\nu}=0
 \rightarrow \,  \varepsilon^{\mu \sigma \delta
\nu}{\partial}_{\sigma} f_{\delta \nu}=0 . \eqno (22)$$ 
It is this equation that reduces on integration, over all $\tau$, to
the two usual homogeneous Maxwell equations.  On the zero mode the $\tau$
derivative disappears and 
therefore Eq.$(21)$ reads for $\mu=0 \,,\,\nu=i$,
$\nabla \times {\bf f}=0$ and for the $\mu=i \, ,\,\nu=j$ components,
$\partial_t{\bf f}+{\bf \nabla}f_0=0$ where we have called $f^{\mu
5}=(f_0,{\bf f})$. Equation $(22)$ then reduces to the
standard homogeneous Maxwell equations; therefore we see that the
homogeneous pre-Maxwell equations ({\it i.e.}, Eq.$(19)$) for the 5 component
fields and the spacetime component fields decouple on the 0-mode.
 This has the effect of reducing
the pre-Maxwell system of equations, as we discuss below as well for
the inhomogeneous equations, to the usual
 Maxwell equations. With appropriate identification of the
integrated quantities, the zero mode of the pre-Maxwell equations
coincides with the Maxwell theory (it is for this reason that the five
dimensional gauge fields associated with the Stueckelberg theory are
called ``pre-Maxwell'' fields).
\par We now turn to the current-dependent pre-Maxwell equations. These
can be written:
$${\partial}_{l_1}n^{l_2 \, l_1}=-j^{l_2}, \eqno(23)$$
where $n^{l_1 l_2}$ are the matter induced (excitation) fields
 (corresponding to ${\bf H}, {\bf D}$ in the 4D theory). 
We remark that, restricting our attention to the spacetime components
of Eq. $(23)$, which then reads
$$ \partial_5 n^{\mu 5} + \partial_\nu n^{\mu \nu} = -j^\mu, \eqno(23')$$
we may extract the 0-mode by integrating over all $\tau$.  Since
$j^k$ satisfies the five dimensional conservation law $\partial_k j^k
=0$, its integral over $\tau$ (assuming$^{12,17}$  $j^5 \rightarrow 0$ for $\tau
\rightarrow \pm \infty$) reduces to the four dimensional conservation
law $\partial_\mu J^\mu =0$, where $J^\mu = \int j^\mu(x,\tau)d\tau$
is the 0-mode part of
$j^\mu$ (this formula for the conserved $J^\mu$ is given in
Jackson$^{22}$). 
 The first term of the left side of $(23')$ vanishes under integration 
(assuming that $n^{\mu 5}\rightarrow 0$ for $\tau \rightarrow \pm
\infty$), and one obtains the form
$$ \partial^\nu F^{\mu \nu} = J^\mu,$$
where we may identify the zero mode fields $F^{\mu \nu}$, as above, with the
Maxwell fields ${\bf H}, {\bf D}$.
  With the zero mode of $(22)$, we
see that the Maxwell theory is properly contained in the five
dimensional generalization we are studying here.  We see that in Eq. $(23)$
the $l_2=\mu$ component reduces on the 0-mode to the standard Maxwell equations
and for $l_2=5$ we get $\partial_t n^0+\nabla\cdot {\bf n}=0$ (using
similar notation as  above, $n^{\mu 5}=(n_0,{\bf n})$); thus for the
pre-Maxwell equation involving the currents, on the 0-mode,
the fields associated with the fifth component and the spacetime fields decouple.
\par We now assume 
the existence of linear constitutive equations in the dynamical
structure of the 5D fields  in a medium 
which connects the $n$ tensor-field to the $k$
tensor-fields using a fifth rank tensor ${\cal E}$ which is a
generalization of the fourth rank covariant permeability- dielectric
tensor$^{23}$ which relates the $E,B$ fields to the excitation fields $D,H$
in the usual Maxwell electrodynamics. The constitutive equations have the form
$$n^{l_1 \, l_2}={\cal E}^{l_1 \, l_2 \, q
_1 \, q_2 \, q_3}k_{q_1 \, q_2 \, q_3}, \eqno (24)$$
antisymmetric in $l_1 l_2$ as well as $q_1 q_2 q_3$ (the indices of
$k$ have been lowered with the generalized Minkowski metric tensor; 
we shall treat other tensors in the same way in the following).
It is useful at this point to  distinguish between the space-time
elements 
 $f_{\mu \nu}$ and the elements $f_{\mu 5}$.  Let us  assume that the
tensor introduced in Eq. $(24)$ does not mix these fields (for
$n^{\mu5}$, if $\varepsilon_{\mu 5 q_1 q_2 q_3}$ has $q_1 q_2 q_3 =
\alpha \beta \gamma$, then the components of $k$ that enter are 
of the form $k_{\alpha \beta \gamma} = {\cal E}_{\alpha \beta \gamma
\mu 5} f^{\mu 5}$ only; similarly, for $n_{\mu \nu}$, only the
components ${\cal E}_{\mu \nu \alpha \beta 5}$ can occur, and
$k^{\alpha \beta 5}$ connects only to the components $f_{\lambda
\sigma}$ of the field tensor). As we have pointed out above, the
vector $f_{\mu 5}$ is physically distinguished from the antisymmetric
tensor $f_{\mu\nu}$ in the inhomogensous field equations, since the
source terms break the higher symmetry of the homogeneous field
equations. The assumption that the constitutive equations do not
couple these components results in a simpler system to analyze,
although (as for Hall type effects in the non-relativistic theory) it is
conceivable that the more general case could occur.
\par We introduce the
new set of fields:
$$b^{\mu \nu}={1 \over 2}\varepsilon^{\mu \nu \lambda
 \sigma}f_{\lambda \sigma},\eqno(25)$$
so that 
$$ n^{\lambda \sigma} = 2 {\cal E}^{\lambda \sigma \alpha \beta
5}b_{\alpha \beta}.\eqno(26)$$
On the zero mode, the fields $b_{\mu \nu}$ correspond to the dual
Maxwell fields; in
this theory they play a role analogous to the ${\bf B}$ fields in the
Maxwell theory. In a similar way, the $f_{\mu5}$ fields are analogous
to ${\bf E}$. The part of the tensor ${\cal E}_{l_1 \, l_2 \, q
_1 \, q_2 \, q_3}$ connecting the $\mu 5$ fields is discussed below.
\par Working with these fields enables us to construct the
equations in a form which, as we shall show, generalizes the Maxwell
theory to a form where the invariant time $\tau$ plays the role of $t$
and spacetime plays the role of space. This analogy helps to interpret
the physics and it distinguishes between the familiar physical
quantites $f_{\mu \nu}$ and the new fields $f_{\mu 5}$.    
Substituting these fields in $(21)$ and $(22)$,
we find 
$${\partial}_5 b^{\mu \nu }+{\partial}_{\sigma} \varepsilon^{\mu \nu
\sigma \lambda }f_{\lambda 5}=0, \eqno (27) $$
analogous to $(1d)$, and
$$ {\partial}_{\sigma} b^{\mu \sigma}=0, \eqno (28)$$
analogous to $(1c)$. For the spacetime excitation fields we define
$$h_{\mu \nu}={1 \over 2}\varepsilon_{\mu \nu \lambda
 \sigma}n^{\lambda \sigma}$$
and we get from $(23)$, for $l_2=\mu$,
$${\partial}_5n^{\mu 5}-{1 \over 2} \varepsilon^{\mu \sigma \lambda
\nu} {\partial}_\sigma h_{\lambda \nu}=-j^\mu, \eqno (29)$$
analogous to $(1b)$,
where we have used
$${\varepsilon_{\alpha \beta \eta \delta }}{ \varepsilon^{\eta \delta
\gamma \mu
}}=-2\bigl({\delta}_\alpha^\gamma{\delta}_\beta^\mu-{\delta}_\alpha^\mu{\delta}_\beta^\gamma
\bigr). \eqno(30)$$ 
To complete the set of equations, we note that for $l_2=5$, we get from $(23)$
$$\partial _\sigma n^{5 \sigma}=-j^5 ,\eqno(31)$$
analogous to $(1a)$.
\par 
 To obtain a mass-energy conservation law for the fields, we multiply
 $(29)$ by $f_{\mu 5}$ and $(27)$ by $h_{\mu \nu}$, and then combine them, obtaining
$$ \bigl[f_{\mu 5} \partial_5 n^{\mu 5}+{1\over 2}h_{\mu \nu}{\partial}_\tau b^{\mu\nu}\bigr]+{1 \over 2} \varepsilon^{\sigma \mu \lambda \nu}{\partial}_\sigma (f_{\mu 5}h_{\lambda \nu})=-j^\mu f_{\mu 5}. \eqno (32)$$
Assuming the dielectric tensor reduced into the $\mu 5$ and $\mu \nu$
subspaces is symmetric (the relations of $n^{\mu 5}$ to $f_{\mu 5}$
and $n^{\mu \nu}$ to $f_{\mu \nu}$ go by the contraction ${\cal
E}\varepsilon$; the exclusive property of indices of $\varepsilon$ then
imply simple conditions on ${\cal E}$ for the symmetry of these forms) we can write $(32)$ as
$${1\over 2}{\partial}_5 \bigl[f_{\mu 5} n^{\mu 5}+{1\over 2}h_{\mu \nu} b^{\mu\nu}\bigr]+ {1 \over 2} \varepsilon^{\sigma \mu \lambda \nu}{\partial}_\sigma (f_{\mu 5}h_{\lambda \nu})=-j^\mu f_{\mu 5}.\eqno (33)$$
From the Stueckelberg Hamiltonian$^{17}$
$$ K = {1 \over 2M} (p^\mu - e_0a^\mu)(p_\mu - e_0a_\mu) - e_0 a_5, \eqno(34)$$
where $e_0$ differs from the usual electric charge by a dimensional
constant, one can derive the relativistic Lorentz force$^{13, 17,21,24}$
$$M{\ddot x}^\mu = e_0 f^\mu\,_\nu {\dot x}^\nu + e_0 f^{\mu 5},
 \eqno(35)$$
from which it follows that
$$ M{\dot x}_\mu{\ddot x}^\mu = M {d \over d\tau}( {\dot x}_\mu{\dot
x}^\mu) = {\dot x}_\mu f^{\mu 5} \eqno(36)$$
\par Since, in the Stueckelberg theory,
the Hamilton equations imply that 
$$ {\dot x}^\mu = {1 \over M}(p^\mu - e_0a^\mu),$$
and hence the $\tau$ derivative in the central equality of $(36)$
corresponds to a change in the mass-squared $(p^\mu -
e_0a^\mu)(p_\mu - e_0a_\mu)$ of the particle. 
It therefore follows that
 $j^\mu f_{\mu 5}$, is the rate of change of the mass of the 
particle.
 We
therefore  identify $s^\sigma=
{1 \over 2}\varepsilon^{\sigma \mu \lambda\nu}f_{\mu 5}h_{\lambda
\nu}$ as the analogue of the Maxwell Poynting vector. This
Poynting 4-vector is the mass radiation of the field.  We see,
furthermore, that ${1 \over 2}\bigl[f_{\mu 5} n^{\mu 5}+{1\over 2}h_{\mu \nu}
b^{\mu\nu}\bigr]$ is the scalar mass density of the field (its four
integral is the dynamical generator of evolution of the
non-interacting field$^{17,19}$).
\par  We now introduce the eikonal approximation, i.e., set 
$$f_{l_1
l_2}(x,\tau)=f_{l_1 l_2}(x)\exp\,i\kappa( \tau-\Psi(x))$$
 for large $\kappa$. 
In the absence of
sources the 5D-Maxwell equations $(27),(28),(29),(31)$ take the form
(for large $\kappa$)
$$ b^{\mu \nu }-\varepsilon^{\mu \nu \sigma \lambda
}p_{\sigma}f_{\lambda 5}=0, \eqno (37) $$
$$ p_{\sigma} b^{\mu \sigma}=0 , \eqno (38)$$
$$n^{\mu 5}+{1 \over 2} \varepsilon^{\mu \sigma \lambda \nu} p_\sigma h_{\lambda \nu}=0, \eqno (39)$$ 
$$p_\sigma n^{5 \sigma}=0, \eqno(40)$$
where $p_\sigma=\partial_\sigma \Psi$.
\par
We now relate the direction of $p_\mu$ to the polarization of the
 fields. We write the ``cross product'' of $n$ and $b$ (analogous to
 the cross product
 of ${\bf D}$ and ${\bf B}$ in Maxwell's theory):
$$\varepsilon_{\mu \nu \sigma \lambda}n^{\nu 5} b^{\sigma \lambda}=2
n^{\nu 5}f_{\nu 5}p_\mu,\eqno(41)$$
or
$$p_\mu={1\over 2 n^{\alpha 5}f_{\alpha 5}}\varepsilon_{\mu \nu \sigma
\lambda}n^{\nu 5} b^{\sigma \lambda},$$
where we have used $(37)$ and $(40)$.
It is clear that since $p_\mu$ and the Poynting four-vector are cross
products of tensors which are not necessarily aligned in the same
four-directions, they are in general not parallel to each other (in
space-time) due to the anisotropy of the medium, i.e., the wave normal
and radiation flow directions are not, in general, the same.
\par The relations $(37)-(40)$, along with the constitutive relations
relating $n^{\mu \nu}, f_{\sigma \lambda}$, and $n^{\mu 5}, f_{\sigma
5}$, provide relations analogous to $(6)$ characterizing the possible
 field strengths of the eikonal
approximation in terms of properties of the medium.  We shall not
treat these relations here, but discuss the mass-radiation
flows, along the rays,  on spacetime
 geodesics in the interesting special case where $h^{\mu \nu}=b^{\mu \nu}$,
which is analogous to the case of materials with $\mu=1$ in Maxwell's
electromagnetism. This case is interesting since, although the space is
empty in the usual sense (in analogy to ${\bf E}={\bf D} \, ,\,
 {\bf B}= {\bf H}$), the  dielectric effect involving the $f_{\mu 5} $ components can
drive the radiation on curved trajectories, i.e., the corresponding
spacetime can have a non-trivial metric structure.  
\par  We multiply $(37)$ by ${1 \over 2}\varepsilon^{\alpha \beta \mu
\nu}p_{\beta}$. We then use $(39)$ and $(30)$ to obtain
$$ {n_\alpha}^5-p_\alpha p^\beta f_{\beta 5}+p_\beta p^\beta
 f_{\alpha 5}=0 \eqno (42)$$
\par Defining the reduced dielectric tensor ${\cal E_\alpha}^\beta$ as the
part of the general dielectric tensor which connects only the $\alpha 5$ components of
the fields, i.e.,
$$n_{\alpha 5}={{\cal E}_\alpha}^\beta f_{\beta 5}, \eqno(43)$$
the condition $(40)$ then implies that ${\cal E}_\alpha^\beta f_{\beta
5}$ cannot be in the direction of $p^\alpha$ (unless it is lightlike).
We  obtain from Eq. $(42)$
$$\bigl({\cal E}_\alpha\,^\beta-p_\sigma p^\sigma {\delta_\alpha}^\beta +p_\alpha p^\beta \bigr)f_{\beta 5}=0, \eqno(44)$$
where we have chosen the negative sign for the signature of the fifth
 index, $n_{\alpha 5}=-{n_\alpha}^5$ [with this choice the flat space
limit, for which ${\cal E}_\alpha\,^\beta = \delta_\alpha\ ^\beta$,
Eq. $(44)$, with $(40)$, admits only spacelike $p_\alpha$; for
positive signature of the fifth index, in this limit, $p_\alpha$ would
be timelike].
\par Eq. $(41)$ has a solution only if the determinant of the coefficients
vanishes (a similar calculation in which the field strengths $f_{\mu
\nu}$ enter in place of $f_{\mu 5}$ results in the same condition on
these coefficients, as it must). It is somewhat simpler to work with
the eigenvalue equation
$(44)$. Assuming as before that this dielectric tensor is symmetric,
we can work
in a Lorentz frame in which it is diagonal.  In this frame we have
(for the transformed fields)
$$ f_{\alpha 5}= -{p_\alpha \over ({\cal E}^\alpha - p^2)}(p^\beta f_{\beta
5}). \eqno(45)$$
Note that in the isotropic case for which all of the ${\cal E}^\alpha$
are equal, one obtains $p_\beta f^{\beta 5}=0$, and
the metric becomes conformal, i.e., one obtains the condition 
$${\cal E}^{-1}\eta^{\mu\nu}p_\mu p_\nu = -1, $$
where $\eta^{\mu\nu}$ is the flat space Minkowski metric $(-1,1,1,1)$.
\par  Multiplying the equation $(45)$
on both sides by $p^\alpha$, and summing over $\alpha$, one obtains
the condition ($p^2 \equiv p_\mu p^\mu$),
$$ 0=K={p_1^2 \over {\cal E}_1-p^2} + {p_2^2 \over {\cal E}_2-p^2} +
{p_3^2 \over {\cal E}_3-p^2} - {p_0^2 \over {\cal E}_0-p^2} +1. \eqno (46)$$
 This condition determines, in this case, the Fresnel surface of the
wave fronts.
\par It then follows that
$$ {\partial K \over \partial p_\mu}={2 p^\mu \over {\cal E}^\mu -p^2}
+2 p^\mu {\partial K \over \partial p^2}. \eqno(47)$$  
 Calculating the  scalar product of $(45)$ and $(47)$ one then obtains
$$\eqalign{f_{\mu 5} &{\partial K \over \partial p_\mu} =\cr
&= -2(p_\nu
f^{\nu 5}){ \{ \sum_{i=1,2,3} {(p^i)^2 \over ({\cal E}^i-p^2)^2}-{{p_0}^2
 \over ({\cal E}^0-p^2)^2} -{\partial K \over \partial p^2}
 \}} = 0  \cr}. \eqno(48) $$
Multiplying the expression $(37)$ for  $b_{\mu \nu} (h_{\mu \nu})$ by
$(47)$, the contribution of the second term of $(47)$
vanishes since the Levi-Civita tensor is antisymmetric; the first
term, according to $(45)$, is proportional to $f^{\alpha 5}$, and
vanishes for the same reason. 
It therefore follows that
$$ {\partial K \over \partial p^\mu}h^{\mu \nu}=0. \eqno(49)$$
Since the scalar product of ${\partial K \over \partial p^\mu}$ with
both $h^{\mu\nu}$ and $f^{\mu 5}$ is zero, it is proportional to their ``cross
product'' i.e., it is parallel to the Poynting vector. To make the
proof explicit, it is convenient to define $V^\mu = {\partial K \over
 \partial p_\mu}$, $H_i = -h_{0j}$, $F_i = f_{i5}$, and $D^i =
\varepsilon ^{ijk}h_{jk}$ (the space index may be raised or lowered
without changing sign in our Minkowski metric).  In this case, the
conditions $V^\mu h_{\mu \nu} =0$ and $V^\mu f_{\mu 5}$ become
$$ \eqalign{V^0{\bf H} - {\bf V} \times {\bf D}&=0 \cr
-V^0 f^{05} + {\bf V}\cdot {\bf f} &=0 \cr
{\bf V} \cdot {\bf H} &=0, \cr} \eqno(50)$$
where we have used boldface to represent the space components of the vector.
In these terms, the Poynting vector is given by
$$ \eqalign{S^0 &=  {\bf D}\cdot {\bf f}\cr
{\bf S} &= f^{05} {\bf D} + {\bf f} \times {\bf H}
\cr} \eqno(51)$$
 \par Taking the cross product of ${\bf f}$ with the first of $(50)$,
one obtains
$$ V^0 ({\bf f} \times {\bf H}) = {\bf V} ({\bf f}\cdot {\bf D}) -
{\bf D} ({\bf f}\cdot {\bf V}).$$
For $V^0 \neq 0$, one may substitute this into the second of
$(51)$.  The $f^{05}{\bf D}$ term, with the help of the second of
$(50)$, cancels, and we are  left with
$$ {\bf S} = {S^0 \over V^0} {\bf V}.$$
It then follows that $S^\mu = {S^0 \over V^0} V^\mu $, i.e., $V^\mu$
is proportional to the Poynting vector.  For the case $V^0 = 0$, the
second and third of $(50)$ imply that 
$$ {\bf V}\cdot {\bf f} = {\bf V} \cdot {\bf H} =0,$$
i.e., if ${\bf V} \neq 0$ (the case $V^\mu =0$ is exceptional in the
eikonal approximation), it must be proportional to ${\bf f} \times
{\bf H}$. From the first of $(50)$, we see that ${\bf V} \times {\bf
D} =0$, and if ${\bf D} \neq 0$, it must be proportional to ${\bf V}$.
The space part of $S^\mu$, from the second of $(51)$ is then
proportional to ${\bf V}$. Under these conditions, the time part of
$S^\mu$ vanishes, and therefore we again obtain the result that
$V^\mu$ is proportional to $S^\mu$. If ${\bf D} = 0$, then $S^0 =0$
and, since ${\bf V}$ is proportional to $ {\bf f} \times {\bf H}$, it
again follows that $V^\mu$ is proportional to $S^\mu$.   
\par From this point, one may follow the same procedure used in the case of
Maxwell's electromagnetism (Eqs.$(14)$ to $(18)$) to obtain the
Hamiltonian flow corresponding to the admissible modes. The Lagrangian
associated with the Hamiltonian $(46)$ satisfies a minimal principle,
from which it follows that the Hamiltonian flow is geodesic on this manifold. 
 Replacing $f^{\beta
5}$ in $(44)$ by $({\cal E}^{-1})^\beta\, _\gamma n^{\gamma 5}$,
one obtains
$$ \bigl\{\delta^\alpha\,_\gamma - M^\alpha\,_{\gamma\mu\nu} p^\mu
p^\nu \bigr\}n^{\gamma 5} = 0, \eqno(52)$$
where
$$ M^\alpha\,_{\gamma\mu}\,^\nu = \delta^\alpha\,_\mu ({\cal
E}^{-1})^\nu\,_\gamma - \delta_\mu\,^\nu  ({\cal
E}^{-1})^\alpha\,_\gamma . \eqno(53)$$
The condition $(40)$ implies that the solutions can lie only in the
hyperplane orthogonal to $p^\sigma$.  The projection of the matrix 
$ M^{\alpha \nu}_{\gamma\mu}p^\mu p_\nu$ (in the indices $\alpha,
\gamma$) into the three dimensional subspace orthogonal to $p^\sigma$
is symmetric, and can therefore be diagonalized by an orthogonal (or
pseudo-orthogonal) transformation in three dimensions.  In fact, the Gauss law
and a gauge condition restrict the polarization degrees of freedom to
three$^{19}$ (the eikonal approximation is far from the zero mode,
which corresponds to the Maxwell limit, for which only two
polarizations survive) and hence one finds three geodesics.
\par For $p^\sigma$ timelike, one can choose a Lorentz frame in which
the eigenvalue condition $(52)$ has the form
$$ \bigl\{\delta ^\alpha\,_\gamma - M'^\alpha_{\gamma 00}(p'^0)^2
\bigr\}n'^{\gamma 5} = 0, \eqno(54)$$   
 and for $p^\sigma$ spacelike,
$$ \bigl\{\delta ^\alpha\,_\gamma - M''^\alpha_{\gamma 33}(p''^3)^2
\bigr\}n''^{\gamma 5} = 0. \eqno(55)$$
 In each of these cases, the matrix can be diagonalized under the little group
acting in the space orthogonal to $p^\sigma$ (leaving it invariant).  For the
 lightlike case, up to a rotation, $p^\sigma$ has the form $(p,0,0, p)$.  The
remaining matrix may then be diagonalized under SO(2) rotations, to
obtain just two geodesics, corresponding to the polarization states of
a massless Maxwell-like theory.  This special limiting case will be
investigated in detail elsewhere.
\par With the same procedure as applied to the Maxwell case treated
above, with $H$ replaced by $K$,
and the space indices replaced by spacetime indices, one finds the
symplectic structure of the flow of matter in space time.
\par It has been shown by Kline and Kay $^2$, as discussed above,
 that for the three
dimensional Maxwell case, the Hamilton equations resulting from the
eikonal coincide with the geodesic flow generated by the resulting
metric (recall that the direction of momentum associated with both
eigenstates is the same); a similar proof can be applied to the 4D
cases we have studied here.
\bigskip
\noindent
{\bf 4. Eikonal of the Relativistic Stueckelberg-Schr\"odinger Equation}
\smallskip 
\par It is interesting to  apply the eikonal method to a relativistic
 Schr\"odinger
equation in a medium which is not isotropic,
for example, in a crystal with shear forces$^{25}$, with locally
varying band structure (as in a crystal under nonuniform stress, or
near the boundaries or impurities).  In this case, the
rays are directly associated with the (probability) flow of particles. The
eikonal eigenvalue condition is one dimensional in this case, since
the field is scalar.  For an analog of this structure (corresponding,
 for example, to
a distribution of events in a potential periodic in both space and
time) 
in four dimensions described by a relativistically covariant
 equation of Stueckelberg-Schr\"odinger type, the metric one obtains is a
spacetime metric, and the geodesic flow is that of
the quantum probability for the spacetime events (matter) described by
the Stueckelberg wave function. We show elsewhere$^{26}$, moreover,
that such an  equation may be derived from a relativistic generalization
of a stochastic procedure
analogous to that of Nelson$^{27}$, with non-trivial correlations
between the stochastic motions in different directions of spacetime.
This simplest analog of the
nonrelativistic problem is given by
$$ i{\partial \over \partial \tau} \psi_\tau (x)= \partial^\mu
{\cal E}_{\mu \nu} \partial^\nu \psi_\tau(x), \eqno(56)$$ 
   where ${\cal E}_{\mu \nu}$ corresponds to the effect of the medium,
and is assumed to be symmetric. 
The Schr\"odinger current is then
$$ j_\tau (x)_\nu = -i \bigl( {\psi_\tau}^* {\cal E}_{\mu \nu}
\partial^\mu \psi_\tau - \psi_\tau{\cal E}_{\mu \nu} \partial^\mu 
{\psi_\tau}^* \bigr) . \eqno(57)$$
In the eikonal approximation, for which the frequency associated with
$\tau$ (essentially the total mass of the particle$^{15}$) is large, 
 one obtains the condition
$$ K = {\cal E}_{\mu \nu} p^\mu p^\nu -1 =0, \eqno(58)$$
analogous to the Fresnel surface condition $(46)$ for the optical
case.  It is clear that  $\partial K/ \partial p_\mu$ is in the direction
of $j^\mu_\tau$. This implies that $K$ is the operator of evolution
for the dynamical flow of particles, corresponding to the rays. It
follows from the Hamilton equations that the
flow is geodesic, where ${\cal E}_{\mu \nu}$ is the metric. 
\bigskip
{\bf 5. Summary and Discussion}
\smallskip
\par We have reviewed the eikonal treatment of Maxwell waves in an
anistropic medium in a form which exhibits the structure of a
canonical mechanics on a three dimensional pseudo-Riemannian manifold,
 where there are two metrics, one
for each admissible polarization mode.  The canonical momenta for each
of the corresponding geodesics have the same direction, but differ in
 magnitude.
\par We then studied the eikonal approximation for the 5D wave equation for the
 generalized Maxwell fields implied by the requirements
of gauge invariance of the Stueckelberg manifestly covariant quantum
theory.  One finds in this case the structure of a
 canonical mechanics on a four dimensional pseudo-Riemannian manifold,
where there are in general three metrics for each of the timelike or
spacelike possibilities for the canonical momenta, and two for the limiting
lightlike case.  No motion of the medium is required.  This canonical 
mechanics has the form of
Stueckelberg's covariant mechanics, where the form of the Hamiltonian
describes the Fresnel surfaces, and the evolution generated by the Hamilton
equations carries mass flow along the
geodesics.
\par We finally studied briefly the eikonal approximation to an non-isotropic
Stueckelberg-Schr\"odinger equation written in a form analogous to the
non-relativistic theory in a crystal as it appears in a 
Brillouin zone, where the local energy surface carries curvature (an
example of a model providing an analogy to a crystal in spacetime is
that of an electromagnetic field in a resonant cavity), or, as derived
from a relativistic Nelson type stochastic procedure with non-trivial
correlations$^{26}$.  The eikonal 
approximation corresponds to (total system) mass
large compared to typical Hamilton-Jacobi momenta. In this case the
Hamiltonian flow is in the direction of the eikonal approximation to
the quantum mechanical current, so that the analog of the Fresnel
surface condition is the generator of evolution for the flow of
particles, corresponding to the rays.  The Hamiltonian flow is
geodesic with respect to the pseudo-Riemannian metric identified with
the effective mass matrix.  This construction provides a model for an
underlying quantum mechanical structure for classical dynamical
motion along geodesics in a pseudo-Riemannian manifold, an analog
to general relativity.
\noindent
{\it Acknowledgements}
\smallskip
 \par We wish to thank C. Piron and F.W. Hehl for
 discussions at an early stage of this work, and H. Goldenberg and N. Erez
 for helpful discussions of propagation of waves in a medium. One of
 us (L.P.H.) wishes to thank S.L. Adler and the Institute for Advanced
 Study for hospitality during the Spring Semester 2003, when this work
 was completed. He is grateful for a grant in aid from the Funds for
 Natural Sciences for this period.

\noindent
{\it References}
\smallskip
\frenchspacing
\item{1.} H. Goldstein, {\it Classical Mechanics}, Addison-Wesley,
N.Y. (1950)
\item{2.}M. Kline and I.W. Kay, {\it Electromagnetic Theory and
Geometrical Optics}, John Wiley and Sons, N.Y. (1965).
\item{3.} P. Piwnicki,{\it Geometrical approach to light in
inhomogeneous media}, gr-qc/0201007.
\item{4.} M. Visser, Class. Quant. Grav. {\bf 15}, 1767 (1998);
L.J. Garay, J.R. Anglin, J.I. Cirac and P. Zoller, Phys. Rev. {\bf
A63}, 026311 (2001); Phys. Rev Lett. {\bf 85}, 4643 (2000).
\item{5.}M. Visser, Carlos Barcelo, and Stefan Liberati, Analogue
models of and for gravity gr-qc/0111111 (2001).
\item{6.} Carlos Barcel\'o and Matt Visser, Int. Jour. Mod. Phys. D {\bf
10}, 799 (2001).
\item{7.} Carlos Barcel\'o, Stefano Liberati and Matt Visser,
Class. and Quantum Grav. {\bf 18}, 3595 (2001).
\item{8.} U. Leonhardt and P. Piwnicki, Phys. Rev A {\bf 60}, 4301
(1999).
\item{9.} V.A. De Lorenci and R. Klippert, Phys. Rev. D {\bf 65},
064027 (2002). 
\item{10.}  Y.N. Obukhov and F.W. Hehl, Phys. Lett. B 458(1999) 466.  See
also, M. Sch\"onberg, Rivista Brasileira de Fisica {\bf 1}, 91 (1971); A Peres,
Ann. Phys.(NY) {\bf19},279 (1962).
\item{11.} H. Urbantke, J. Math. Phys. {\bf 25} 2321(1984); Acta Phys.
 Austriaca Suppl. XIX,875 (1978).

\item{12.} E.C.G. Stueckelberg, Helv. Phys. Acta {\bf 14}, 322, 588 (1941);
J.S. Schwinger, Phys. Rev. {\bf 82}, 664 (1951); R.P. Feynman,
Rev. Mod. Phys. {\bf 20}, 367 (1948); R.P. Feynman, Phys. Rev. {\bf 80},
 440 (1950). C. Piron and L.P. Horwitz, Helv. Phys. Acta {\bf 46}, 316
(1973), extended this theory to the many-body case throught he
postulate of a universal invariant evolution parameter.
\item{13.} O. Oron and L.P. Horwitz, Phys. Lett. {\bf A 280}, 265 (2001).
\item{14.} P.A.M. Dirac, Proc. Roy. Soc. London, Ser. A {\bf 167}, 148
(1938).
\item{15.} R.I. Arshansky and L.P. Horwitz, Jour. Math. Phys. {\bf
30}, 66 and 380 (1989).
\item{16.} M.A. Trump and W.C. Schieve, {\it Classical Relativistic
Many-Body Dynamics}, Fund. Theories of Physics, Kluwer, Dordrecht
(1999).
\item{17.} D. Saad, L.P. Horwitz and R.I. Arshansky, Found. Phys. {\bf
19}, 1126 (1989).  
\item{18.} M.C. Land, N. Shnerb and L.P. Horwitz, Jour. Math. Phys. {\bf 36},
3263 (1995).
\item{19.} N. Shnerb and L.P. Horwitz, Phys. Rev {\bf A48}, 4058
(1993).
\item{20.} M. Visser, {\it Birefringence versus bi-metricity},
contribution to Festschrift in honor of Mario Novello (2002).
\item{21.} M.C. Land and L.P. Horwitz, Found. Phys. Lett. {\bf 4}, 61
(1991).
\item{22.} J.D. Jackson, {\it Classical Electrodynamics}, 2nd edition,
Wiley, N.Y. (1975).
\item{23.} A.O. Barut, {\it Electrodynamics and Classical Theory of
Fields and Particles}, Dover, N.Y. (1964).
\item{24.} C. Piron (personal communication) has obtained this result 
independently.
\item{25.} H. Brooks, Adv. in Electronics {\bf 7}, 85 (1955);
C. Kittel, {\it Quantum Theory of Solids}, p. 131, John Wiley and
Sons, N.Y. (1963).
\item{26.} O. Oron and L.P. Horwitz, in preparation.
\item{27.}  Edward Nelson, {\it Dynamical Theories of Brownian Motion},
 Princeton University Press, Princeton (1967); Edward Nelson, {\it Quantum
Fluctuations}, Princeton University Press Princeton (1985). See also, Ph. Blanchard, Ph. Combe and W. Zheng,
{\it Mathematical and Physical Aspects of Stochastic Mechanics},
Springer-Verlag, Heidelberg (1987), for further helpful discussion.

\vfill
\end
\bye